\documentclass[a4paper,12pt]{article}

\usepackage{graphicx}
\usepackage{amssymb}

\begin{document}
\vspace*{-.6in} \thispagestyle{empty}
\begin{flushright}
\end{flushright}
\baselineskip = 20pt

\vspace{.5in} {\Large
\begin{center}
{\bf Gravitino Dark Matter in Gravity Mediation}

\end{center}}

\vspace{.5in}

\begin{center}
{\bf  J\"orn Kersten$^{a}$ and 
  Oleg Lebedev$^{b} $}  \\

\vspace{.5in}

$^a$\emph{University of Hamburg, II.~Institute for Theoretical Physics\\
Luruper Chaussee 149, 22761 Hamburg, Germany }
\\[3mm]

$^b$\emph{DESY Theory Group\\ 
Notkestra\ss{}e 85, 22603 Hamburg, Germany }
\end{center}

\vspace{.5in}

\begin{abstract}
We study general conditions for  the gravitino to be  the lightest supersymmetric 
particle (LSP) in models with gravity mediated supersymmetry breaking.
We find that the decisive quantities are the K\"ahler potential $K$  and
the gauge kinetic function $f$. In constrained MSSM (CMSSM)  type
models, the gravitino LSP occurs if
the gaugino mass at the GUT scale is greater than approximately 2.5 gravitino masses.
This translates into
$\sqrt{K''}/f' < 0.2 $, where the derivatives are taken with respect to the dominant SUSY breaking field. This requirement can easily be satisfied  in string-motivated 
 setups.
\end{abstract}

\noindent

\newpage

\section{Introduction}

It is an exciting possibility that dark matter has a supersymmetric origin.
Various species can have the properties of dark matter depending on the 
supersymmetry breaking mechanism and further particulars of the model,
with the neutralino and gravitino being the most prominent candidates. 

In classes of models like anomaly \cite{Randall:1998uk}  and mirage \cite{Choi:2005ge}  mediation,
the gravitino is heavier than the other sparticles  and  thus 
cannot constitute dark matter. On the other hand, in gauge 
\cite{Dine:1993yw}  and gaugino \cite{Kaplan:1999ac}  mediation,
the gravitino is light and  represents  a  good candidate for dark
matter \cite{Dine:1993yw},\cite{Buchmuller:2005rt}. 
In gravity mediation \cite{Nilles:1982ik}, the situation is more model-dependent  and both relatively heavy and light
gravitinos are possible. In this work, we study the circumstances under which the gravitino
is the LSP in gravity mediation. If we further require  R- or matter parity 
\cite{Dimopoulos:1981dw}, which can descend from string theory
\cite{Lebedev:2007hv}, the gravitino is stable  and can constitute dark
matter.
Phenomenology of the gravitino LSP has been an active research 
subject \cite{Moroi:1993mb}--\cite{Brandenburg:2005he},
while in this paper we focus on its  supergravity side and identify relevant
constraints on fundamental supergravity quantities.

\section{Supergravity preliminaries}

Let us review relevant  features of the supergravity formalism,
following \cite{Brignole:1997dp} (and the original work \cite{Soni:1983rm}). 
The supergravity scalar potential is given (in Planck units)  by  
\begin{equation}
V=\bar F^{\bar i} F^j K_{\bar i j} -3 ~{\rm e}^G ~.
\end{equation}
Here the subscript $ l $ ($\bar l$)  denotes differentiation with respect to 
the $l$-th ($l$-th complex conjugate)  scalar field.  
$G$ is a function of the K\"ahler potential $K$ and the superpotential $W$,
$G=K+\ln(|W|^2)$, and the SUSY breaking $F$-terms are 
$F^i= {\rm e}^{G/2} K^{i \bar j} G_{\bar j}  $ with $K^{i \bar j} $ being
the inverse of $K_{\bar i j}$. The gravitino mass is given by
\begin{equation}
m_{3/2} = {\rm e}^{G/2} \;. 
\end{equation}
Another quantity we need is the K\"ahler metric $ \tilde K_\alpha$   for the observable fields
$\phi^\alpha$. It is found by expanding the K\"ahler potential around
$\phi^\alpha =0$\,:
\begin{equation}
 K   =  K\Big\vert_{\phi_\alpha =0}  + \tilde K_\alpha \phi^{*\bar\alpha}  \phi^{\alpha}
+ ...
\end{equation}
Then the soft SUSY breaking terms are given by  \cite{Brignole:1997dp}
\begin{eqnarray}
\label{softterms}
 M_a &=& {1\over 2}({\rm Re}f_a)^{-1} F^m \partial_m f_a\;,  \\
 m_\alpha^2 &=& m^2_{3/2} - \bar F^{\bar m} F^n \partial_{\bar m} \partial_n
\ln \tilde K_\alpha\;,\nonumber\\
 A_{\alpha \beta \gamma} &=& F^m \left[ K_m +\partial_m \ln Y_{\alpha \beta \gamma}
-\partial_m \ln (\tilde K_\alpha \tilde K_\beta \tilde K_\gamma)
\right]\;,\nonumber
\end{eqnarray}
where $f_a$ are the gauge kinetic functions,
\begin{equation}
{\rm Re} f_a ={1\over g^2_a}~,
\end{equation}
 $\partial_m \equiv \partial/\partial \phi^m$, and $ Y_{\alpha \beta \gamma}$ are the superpotential
Yukawa couplings.

Vanishing of the vacuum energy requires
\begin{equation}
\bar F^{\bar i} F^j K_{\bar i j} = 3 \, m_{3/2}^2
\end{equation}
at the minimum of the scalar potential.
Thus, the magnitude of the $F$-terms depends on the K\"ahler metric of the SUSY breaking fields
$K_{\bar i j}$. The $F$-terms and, consequently, the soft masses can
be  much larger than the gravitino mass provided the  K\"ahler metric
is sufficiently small.

\section{Gravitino LSP}

Let us consider the case of a single dominant SUSY breaking field $C$.
Omitting for simplicity complex phases, we have 
\begin{equation}
F= \sqrt{3\over K''}~ m_{3/2} \;,
\end{equation}
where $K'' \equiv K_{\bar C C}$. 
Let us further assume universal gauge kinetic functions and K\"ahler metrics for the matter fields.  Then  the soft terms simplify to
\begin{eqnarray}
m_{1/2} &=& \sqrt{3\over 4 K''} ~g^2 f' ~m_{3/2} ~, \nonumber\\
m_0^2 &=& \biggl[ 1- 3 {  \tilde K'' \tilde K - {\tilde K}^{'2} \over K'' \tilde K^2   }   \biggr]~
m_{3/2}^2 ~,
\nonumber\\
A_0 &=& \sqrt{3\over  K''}~ \biggl[ K' -3 {\tilde K' \over \tilde K}   \biggr]~m_{3/2}~,
\label{SoftMassesSingleField}
\end{eqnarray}
where a prime denotes differentiation with respect to $C$ and a double prime
differentiation with respect to $C$ and $\bar C$. We have assumed that the Yukawa couplings are
independent of $C$, which allows us to avoid strong constraints from
electric dipole moments \cite{Abel:2001cv}.

This case corresponds to the CMSSM\@.
We see that the scalar and gaugino masses can be made arbitrarily large by decreasing
$K''$. This requires a non-negligible $f'$ and  
$\tilde K'' \tilde K - {\tilde K}^{'2}
\leq 0 $. 
If this quantity vanishes at the GUT scale, large scalar masses are induced by the
renormalization group evolution down to the electroweak (EW)  scale. Since the gravitino 
mass does not run, we have 
\begin{equation} 
M_i \; , \; m_\alpha \,>\, m_{3/2} 
\end{equation}
and the gravitino is the LSP\@. The $A$-terms usually do not play any significant
role unless they are much larger than the other soft parameters.

Let us quantify this effect. In the CMSSM, the lightest superparticle is either a neutralino
(mostly bino)
or a stau. Unless $\tan\beta$ is large, their masses can be approximated by
 \cite{Cerdeno:2005eu},\cite{Martin:1993ft} 
\begin{eqnarray}
 m_\chi &\simeq& 0.4 \, m_{1/2} \;,\nonumber\\
 m^2_{\tilde \tau} &\simeq& m_0^2 + 0.15 \, m_{1/2}^2 \;, \label{masses}
\end{eqnarray}
where we have neglected the EW contributions. 
For moderate $m_0$, the lightest neutralino is lighter than the staus.
Then the gravitino is the LSP for
\begin{equation}
m_{1/2} > 2.5 \, m_{3/2} \;. \label{2.5}
\end{equation}
For small $m_0$, the stau is lighter than the neutralino,
so that a larger $m_{1/2}$ is required for a gravitino LSP\@.
However, Eq.~(\ref{masses}) shows that the change in the lower bound
will be rather small, so that Eq.~(\ref{2.5}) is still a good
approximation.

Thus, the key parameter is the gaugino mass $m_{1/2}$ and as long 
as the scalar masses squared are non-negative\footnote{One may in principle allow tachyonic
scalar masses at the GUT scale as long as they evolve to positive values at the EW scale
and the EW vacuum is sufficiently long lived \cite{Riotto:1995am}. In this case, there are many deep color and charge breaking vacua, however the EW vacuum is preferred cosmologically \cite{Riotto:1996xd},\cite{Ellis:2008mc}.  }, we obtain the gravitino LSP for
\begin{equation}
{\sqrt{K''}\over f' } < 0.2 \;, \label{bound}
\end{equation}
where we have used $g^2(M_{\rm GUT}) \simeq 1/2 $. This bound  does not involve the superpotential
nor the K\"ahler metric for the observable fields and is therefore 
largely model-independent.

\begin{figure}
\includegraphics{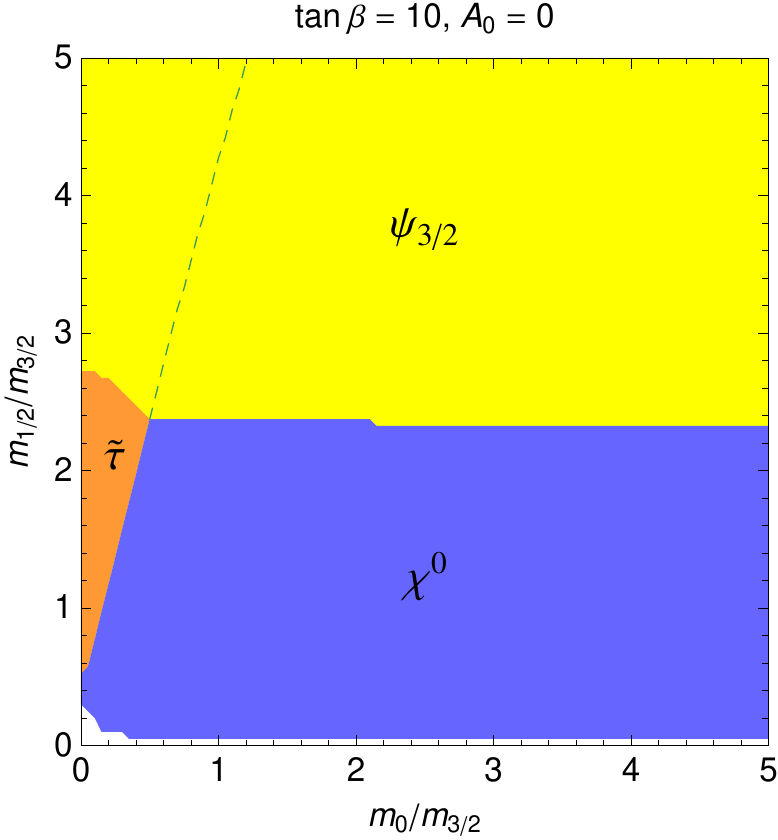}
\hfill
\includegraphics{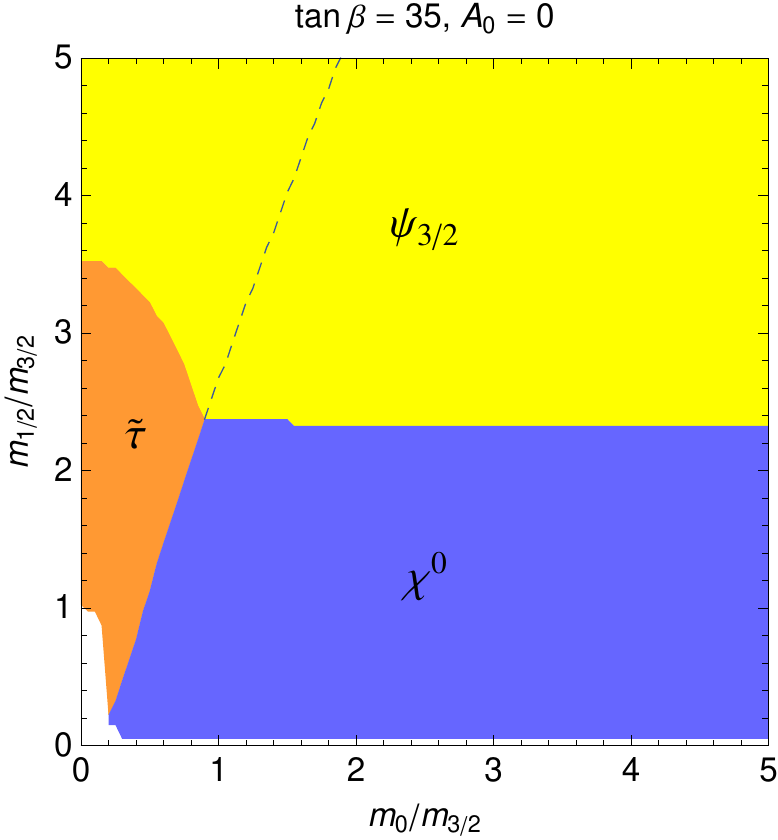}
\caption{Regions with different LSPs in the CMSSM\@.  In the region to the
left of the dashed
line, the stau is the next-to-LSP (NLSP), while to the right the
lightest neutralino is the NLSP\@.  The white area at the bottom of each
plot is excluded because of tachyons or no EW symmetry breaking.
For definiteness, we have fixed $m_{3/2}=400$~GeV and $\mu>0$.}
\label{fig1}
\end{figure}

\begin{figure}
\centering
\includegraphics{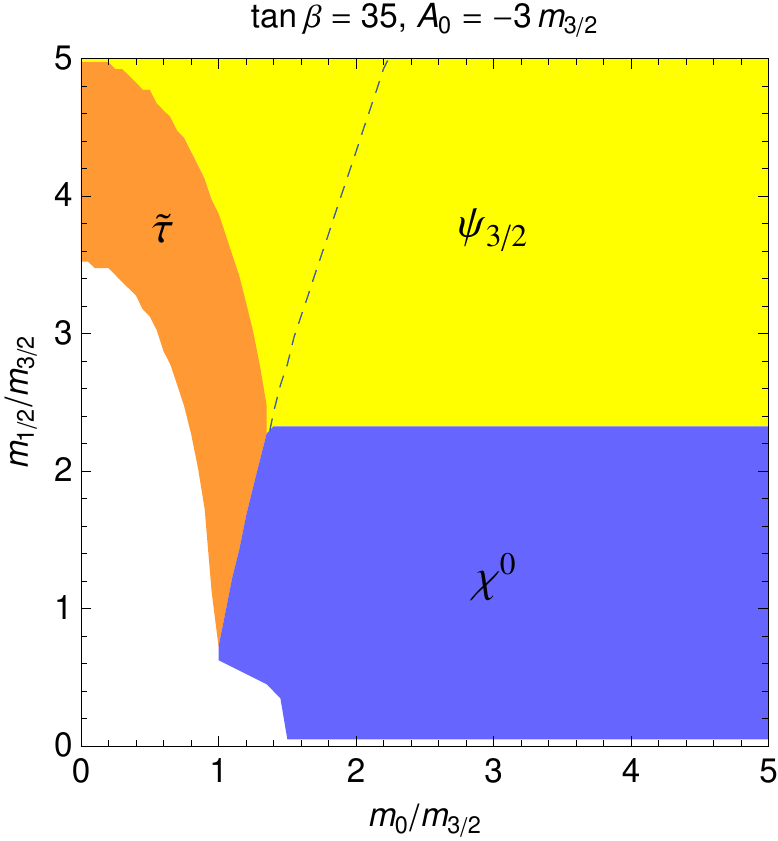}
\caption{Same as Fig.~\ref{fig1}, but with a large $A$-term.}
\label{fig2}
\end{figure}

In Figs.~\ref{fig1} and \ref{fig2} we illustrate these results with a
numerical analysis in the CMSSM\@.  We have used SOFTSUSY 
\cite{Allanach:2001kg} to determine the low-energy superpartner
spectrum.
The figures display parameter space  regions with the gravitino,
neutralino and stau LSP\@.
We have chosen $m_{3/2}=400$~GeV to fix the overall
mass scale for definiteness, while the qualitative
features of the plots are independent of this.

At large $\tan\beta$, Eq.~(\ref{masses}) receives corrections mainly from the $\tau$ Yukawa coupling
so that the $\tilde \tau$ becomes the LSP at small $m_0$. This is displayed
in the right panel of Fig.~\ref{fig1}.
Also, large $A$-terms further decrease the stau mass and can lead to
tachyons, so that the
$\tilde \tau$ LSP and the excluded regions are enlarged, cf.\
Fig.~\ref{fig2}.

Let us  note that our parameter space is subject to  further 
 (model-dependent)  
phenomenological constraints. 
These depend on the overall mass scale,  assumptions on the cosmological history,
whether R-symmetry is exact  or approximate and whether the gravitino constitutes
all or just part of the observed dark matter. 
For instance, as the LEP bound on the Higgs mass requires 
$m_{1/2} \gtrsim 400$~GeV, Eq.~(\ref{SoftMassesSingleField})
implies $m_{3/2} \gtrsim 10^2$~GeV for 
$\sqrt{K''} = \mathcal{O}(10^{-1})$ and $f' = \mathcal{O}(1)$.
For detailed studies, we refer
an interested reader to Refs.~\cite{Moroi:1993mb}--\cite{Brandenburg:2005he}.

From Eq.~(\ref{bound}),
we see  that $K''$ is not required to be very small. In fact, it is of the order of magnitude 
of typical $K''$ expected in string theory. The usual moduli/dilaton  K\"ahler potential is
$K=-a \ln (C + \bar C)$  with $a= {\cal O}(1)$ so that
\begin{equation}
\label{T}
\sqrt{K''} =  { \sqrt{a} \over C+\bar C  }\;. 
\end{equation}
Just to have an idea of the numerics, take $C$ to be the dilaton of the heterotic string. 
Then  $a=1$, $f=C$ and  $C=2$ at the minimum of the potential, 
as required by the observed values of the gauge couplings.  We get
${\sqrt{K''}\over f' } = 0.25$, 
which falls just short of the bound (\ref{bound}).
One should keep in mind, however, that the dilaton-dominated SUSY breaking is not possible 
with the above K\"ahler potential and in realistic cases one has to  include either
non-perturbative corrections to the K\"ahler potential 
\cite{Binetruy:1996xja},\cite{Casas:1996zi} or additional fields.
In the former case, one typically has $K'' \ll 1$ at the minimum of the potential \cite{Barreiro:1997rp},\cite{Buchmuller:2004xr}, although
the zero vacuum energy is not enforced.

When $C$ is the modulus associated with the radius of the compact dimensions,
one can trust the supergravity approximation for $C \gg 1$ in which case $\sqrt{K''}$
can be arbitrarily  small. In realistic cases, however, one has to include more than one field
to have the correct gauge coupling, e.g. $f= C_1 + C_2$. Otherwise, the gauge coupling 
becomes too small.   

One can also entertain the possibility that $C$ is a hidden matter-like field with the 
K\"ahler potential $K= {\rm const} +  \kappa \bar C C $, which dominates SUSY breaking \cite{Lebedev:2006qq}. 
In this case, $\kappa$ can be very small due to large moduli,
$\kappa= 1/(T +\bar T)^n$. 
However, the gauge kinetic function is then given predominantly by some
other field, e.g.\ the dilaton, so that  $f' \ll 1$. 
Consequently,
$\sqrt{K''}/f'$
can be sufficiently small for the gravitino to be the LSP, yet
$\sqrt{K''}/f' \ll 1$ would require careful engineering.

The above formulae can be generalized to the case of multiple SUSY breaking fields 
in a straightforward manner.

\subsection{Semi-realistic example}
 
Let us illustrate with an example how the gravitino LSP
can arise in string-motivated setups.
Consider two modulus-type fields $C_1$ and $C_2$ with
\begin{eqnarray} 
&& K^{\rm hid} =-a \ln(C_1 +\bar C_1) -b \ln (C_2 +\bar C_2) ~, \nonumber\\
&& \tilde K = (C_1 + \bar C_1)^n ~, \nonumber\\
&& f= C_1 - C_2 \;,
\end{eqnarray}
and  $a,b >0$.
Here $n$ is the matter ``modular weight'', which can be negative, positive or zero 
\cite{Ibanez:1992hc}.
For $a+b>3$, locally stable vacua with zero (or small) vacuum energy are possible
\cite{GomezReino:2006dk}. For instance, this is the case when $C_{1,2}$ are 
 the overall modulus and the dilaton.   
Finally, one can choose an appropriate superpotential such that the fields stabilize
at 
\begin{equation}
(C_1 - C_2) \vert_{\rm min} \simeq 2 \;, 
\end{equation}
in order to have the correct gauge couplings at the GUT scale. 
``Mixed'' gauge kinetic functions of this type appear in string models with fluxes
\cite{Lust:2004cx},\cite{Abe:2005rx}.

Requiring zero vacuum energy at the minimum of the scalar potential and neglecting complex phases, we can
parametrize  supersymmetry breaking  by the Goldstino angle $\theta$:
\begin{equation}
F_1=  \sqrt{3\over K^{\rm hid}_{C_1 \bar C_1}}~ m_{3/2}~ \cos \theta ~~,~~
F_2=  \sqrt{3\over K^{\rm hid}_{C_2 \bar C_2}}~ m_{3/2}~ \sin \theta ~~.
\end{equation}
Then the soft terms read
\begin{eqnarray}
m_{1/2} &=& \sqrt{3\over 4 } ~g^2 \Biggl[  {C_1 + \bar C_1 \over \sqrt{a}} ~\cos\theta -
{C_2 + \bar C_2 \over \sqrt{b}} ~\sin\theta  \Biggr]   ~m_{3/2} ~, \nonumber\\
m_0^2 &=& \biggl[ 1+  { 3n \over a    } ~\cos^2\theta  \biggr]~
m_{3/2}^2 ~,
\nonumber\\
A_0 &=& -\sqrt{3}~ \biggl[ \sqrt{a} \left(1+{ 3n \over a    } \right) ~\cos\theta +
\sqrt{b }~ \sin\theta    \biggr]~m_{3/2}~.
\end{eqnarray}
For a sufficiently large $C_1$ and $C_2= C_1-2$, the gauginos are heavy 
so that $\vert m_{1/2} \vert > 2.5 \, m_{3/2}$ and the gravitino is the
LSP\@.
Note that the $C_i$ dependence cancels out in $m_0^2$ and $A_0$.

Taking as an example $C_{1,2}$ to be the overall modulus and the dilaton,  
$a=3$ and $b=1$. Then for $\cos\theta \sim 1$, the gravitino LSP imposes 
the bound $C_1> 5 $. Note that the scalar masses are non-tachyonic at the 
GUT scale for  $n \geq -1$.

Let us finally note that the superpotential does not play a role in this discussion
as long as it stabilizes the fields at the desired values with vanishing vacuum energy.
This question can be analyzed locally, in terms of 
$\delta C_1 \equiv C_1 - C_1 \vert_{\rm min}$ and $\delta C_2 \equiv C_2 - C_2 \vert_{\rm min}$,
 along the lines of Ref.~\cite{Lebedev:2006qq}.
As a result, the superpotential expansion coefficients $x_a$, $W=x_0 + x_1 ~\delta C_1 + x_2~ \delta C_2
+ x_{11}~ (\delta C_1)^2 + x_{12}~ \delta C_1 \delta C_2 + x_{22}~ (\delta C_2)^2+...$,
have to satisfy certain (model-dependent)  constraints.

\section{Conclusions}

We have studied the conditions for the gravitino to be the LSP in
models of gravity mediated SUSY breaking.  
This requirement constrains mainly the gaugino mass parameter at the GUT scale
while the other parameters play a minor role, as long as
the scalar masses are non-tachyonic.
For CMSSM-type models at moderate $\tan\beta$,
the resulting constraint on the K\"ahler potential and the gauge kinetic function is 
approximately   
$\sqrt{K''}/f' < 0.2 $, where the derivatives are taken with respect to the dominant SUSY breaking field.
For large $\tan\beta$ and $A$-terms, the above constraint gets modified at small values
of the universal scalar mass $m_0$. The results are illustrated in
Figs.~\ref{fig1} and \ref{fig2}.

The condition $\sqrt{K''}/f' < 0.2 $ can  easily  be satisfied in string-motivated 
set-ups and thus the gravitino LSP is a reasonable  alternative to the 
neutralino LSP in gravity mediation. 
As it is hard to obtain an extremely small value of $K''$
without finetuning, the gravitino mass is still expected to be of the
same order of magnitude as the other soft masses.

\end{document}